\documentclass[12pt]{article}

\usepackage[dvips]{graphics}

\title{Quantum pattern formation dynamics of photoinduced nucleation process}
\author{Kunio Ishida\\
Corporate Research and Development Center, \\
Toshiba Corporation\\
1 Komukaitoshiba-cho, Saiwai-ku, Kawasaki 212-8582,Japan\\
and\\
Keiichiro Nasu\\
Solid State Theory Division, \\
Institute of Materials Structure Science, KEK,\\
Graduate University for Advanced Study, and CREST JST\\
1-1 Oho, Tsukuba, Ibaraki 305-0801, Japan}

\begin{document}

\maketitle 

\begin{abstract}
We study the dynamics of quantum pattern formation processes in molecular crystals which is a concomitant with photoinduced
nucleation.
Since the nucleation process in coherent regime is driven by the nonadiabatic transition in each molecule followed by the propagation of
phonons, it is necessary to take into account the quantum nature of both electrons and phonons in order to pursue the dynamics of the 
system.
Therefore, we employ a model of localized electrons coupled with a quantized phonon mode and solved the time-dependent Schr\"odinger equation numerically.

We found that there is a minimal size of clusters of excited molecules which triggers
the photoinduced nucleation process, i.e., nucleation does not take place unless sufficient photoexcitation energy
is concentrated within a narrow area of the system.
We show that this result means that the spatial distribution of photoexcited molecules plays an important role
in the nonlinearity of the dynamics and also of the optical properties observed in experiments.
We calculated the conversion ratio, the nucleation rate, and correlation functions to reveal the 
dynamical properties of the pattern formation process, and the initial dynamics of the photoinduced
structural change is discussed from the viewpoint of pattern formation.
\end{abstract}

\maketitle

\section{Introduction}
\label{intro}

Recently coherent control quantum-mechanical states of materials with 
arbitrarily designed optical pulses has been of interest\cite{cc}, 
and the handling of the quantum-mechanical states is expected to lead to novel device applications
in quantum information technology\cite{qi}, for example.
In order to realize and establish such control methods or devices, it is required to reveal the dynamical
properties of quantum-mechanical states in coherent regime.

On the other hand, it was also found that injection of photoexcited states 
induces cooperative phenomena regarding the change of structural, magnetic, or
ferroelectric properties\cite{ct,pda,spin,binuclear,letard}.
These photoinduced cooperative phenomena in condensed matter have
brought us fruitful theoretical and experimental problems with nonequilibrium dynamics of excited states, and
many studies have been presented to clarify their mechanism\cite{nasu,ogawa,nasu2,ishida1,ishida2,ishida3}.
In particular, coherent dynamics of the photoinduced cooperative phenomena is relevant to control such
cooperativity by designed optical pulses.

It has been pointed out that two important situations should be considered to understand the mechanism of photoinduced cooperativity.
In the first one, photoexcited electrons are itinerant and 
long-range interaction between atoms is induced by them.
As a result macroscopic structural change takes place following the appearance of lattice instability\cite{bismuth}.
The second one, which we focus on in this paper, concerns the excitation of localized electrons in each molecular unit followed by the nucleation process due to intermolecular interactions.
In this case the nucleation process is driven by energy transfer
process between molecular units through excitation of vibrational modes (phonons), and the subtle balance between electronic-vibrational excitations dominates the dynamical properties of the atomic motion and the electronic states.
We also note that, in the very initial stage of nucleation process, the
coherent nature of the quantum-mechanical states of electrons and phonons
plays a dominant role, and that the nonadiabaticity of the electrons and phonons should be taken into account
to pursue the dynamics of the system.

Since the pioneering works by Landau\cite{landau} and
Zener\cite{zener}, dynamics of nonadiabatic transitions has been
extensively studied.
We mention that the bifurcation rate of wavefunction was analytically obtained in
general cases\cite{zhu}, and that the wavefunctions before/after
nonadiabatic transition have been understood. 
However, the dynamics of quantum-mechanical states during nonadiabatic transition is
important in, for example,  photochemical reactions\cite{chemical},
and hence computational methods of the dynamics have been proposed by many authors\cite{calc1}.
Since the atomic degrees of freedom are treated as classical variables
in those methods, they are limited to discuss the wavefunctions after decoherence of these variables takes place.
Since, however, the dynamical properties of photoinduced phenomena is
important to discuss the possibility coherent control of
them, it is required to study the coherent dynamics of relaxation
processes which lead to the macroscopic structural change, for example.
Therefore, in order to understand the dynamics of photoinduced nucleation process in coherent regime, we proposed a model of 
electron-phonon systems in which nonadiabatic transition is calculated rigorously by 
quantizing the relevant vibrational modes\cite{ishida1}.

The initial dynamics of photoinduced nucleation process also involves nonlinear pattern formation.
Dynamics of nonlinear spatio-temporal pattern formation has been studied to understand the various aspects of nonequilibrium phenomena
such as the phase separation dynamics in the kinetic Ising model\cite{ising}.
It has been pointed out that the density fluctuation of
relevant physical properties is a ``seed'' of growing patterns,
and that the initial density distribution determines their complicated structure.
Hence, in analogy to them, it is required to study the domain growth dynamics in the presence of
excitation density fluctuation in order to clarify and understand the
nonlinear nature of the photoinduced cooperativity.
However, the previous results on pattern formation process are based on stochastic simulations which describe the kinetics of the system,
and do not correspond to the dynamics of wavefunctions in coherent regime, i.e., nonadiabaticity of the 
electrons and phonons cannot be taken into account by stochastic simulations.
Although the effect of decoherence eliminates such coherent properties within a few picoseconds after
photoexcitation, the very first process of photoinduced cooperativity should be discussed in an ideal situation,
i.e., in coherent regime.

In this paper, we theoretically study quantum pattern formation processes
based on a model of molecular crystals.
We focus on the deterministic solution of the time-dependent Schr\"odinger equation
 in order to clarify and understand the nonlinear nature of the photoinduced cooperativity.
In our previous paper\cite{ishida2}, we showed that the the dynamics of
 pattern formation and the nonlinear dependence of the conversion ratio
are understood with the above-mentioned model.
We present our calculated results on the initial dynamics of the pattern formation in photoinduced
nucleation process in detail.

The organization of the paper is as follows: in Sec.\ \ref{models}
the molecular model is introduced and the method of calculation is
described.
In Sec.\ \ref{results} the calculated results are shown. Sec.\
\ref{summary}
is devoted to summary and conclusions.

\section{models and method}
\label{models}

As we discussed in our previous papers\cite{ishida1,ishida2},
nonadiabatic transitions between quantized states are 
important to study the dynamical aspects of photoinduced cooperative phenomena.

In this paper, we focus on the initial dynamics of a photoexcited state
in interacting molecules, fully quantizing the relevant vibration modes.
However, the dimension of the Hilbert space for the whole system
increases drastically by quantizing atomic variables, which means that numerical
calculation on those systems requires lots of computational
resources.
Thus, we employ a simplest model which is sufficient to describe the
photoinduced nucleation processes.

First, we consider molecules arrayed on a square lattice.
All the electrons in the system are assumed to be localized in each molecule, and only two electronic levels coupled
with a single vibration mode are taken into account per molecule.
Hence the nucleation processes occur not through electron transfer but
through excitation energy transfer between molecules.
Each molecule has crossing diabatic PESs which are relevant to the nonadiabatic
transitions mentioned in the last section.
The nonadiabaticity in the dynamics is taken into account via ``spin-flip''
interaction between two electronic states, {\it i.e.}, the interaction strength
between two electronic levels are assumed to be a constant.
Furthermore, we apply the harmonic approximation to the vibration modes.
We mention that this model is known as a model to discuss the relaxation dynamics of, 
{\it e.g.}, photoisomerization of molecules and studied by many authors\cite{mol}.

As for the intermolecular interaction, we take into account three types
of intermolecular interactions described below:
\begin{enumerate}
\item Interaction between the vibration mode of different molecules. It is also responsible for the
      dispersion of the relevant phonon mode.
\item dipole-dipole interaction between excited molecules. The interaction strength is taken up to
      the first order of the molecular distortions.
\item Electron-vibrational interaction which describes the distortion of  molecules induced by the
    excited-state dipole in the adjacent ones.
\end{enumerate}

Hence, the Hamiltonian in the present study is described by:
\begin{eqnarray}
{\cal H} & = & \sum_{\vec{r}} \left \{\frac{p_{\vec{r}}^2}{2}+ \frac{\omega^2 u_{\vec{r}}^2}{2}+ ( \sqrt{2\hbar \omega^3}su_{\vec{r}}+ \varepsilon \hbar \omega + s^2 \hbar \omega ) \hat{n}_{\vec{r}}+\lambda \sigma_x^{\vec{r}} \right \}\nonumber \\
& + &\sum_{\langle \vec{r},\vec{r'}\rangle} [ \alpha \omega^2
     (u_{\vec{r}}-\beta\hat{n}_{\vec{r}})(u_{\vec{r'}}-\beta\hat{n}_{\vec{r'}}) - \{ V - W (u_{\vec{r}}+u_{\vec{r'}}) \} \hat{n}_{\vec{r}} \hat{n}_{\vec{r'}} ],
\label{ham}
\end{eqnarray}
where $p_{\vec{r}}$ and $u_{\vec{r}}$ are the momentum and coordinate
operators for the vibration mode of a molecule at site $\vec{r}$, respectively.
The electronic states at site $\vec{r}$ are denoted by $|\downarrow \rangle_{\vec{r}}$ (ground
state) and $|\uparrow \rangle_{\vec{r}}$ (excited state) and
$\sigma_i^{\vec{r}}\ (i=x,y,z)$ are the Pauli matrices which act only on
the electronic states of the molecule at site $\vec{r}$.
$\hat{n}_{\vec{r}}$ denotes the density of the electron in $|\uparrow
\rangle_{\vec{r}}$ which is rewritten as $\hat{n}_{\vec{r}}=\sigma_z^{\vec{r}}+1/2$.
The second sum which gives the intermolecular interaction is taken over
all the pairs on nearest neighbor sites, where the Coulomb interaction
between excited state electrons are modified by molecular distortion.
The vibrational period of an individual molecule is denoted by $T=2\pi/\omega$ in
the rest of the paper.

\begin{figure}[h]
\scalebox{0.8}{\includegraphics*{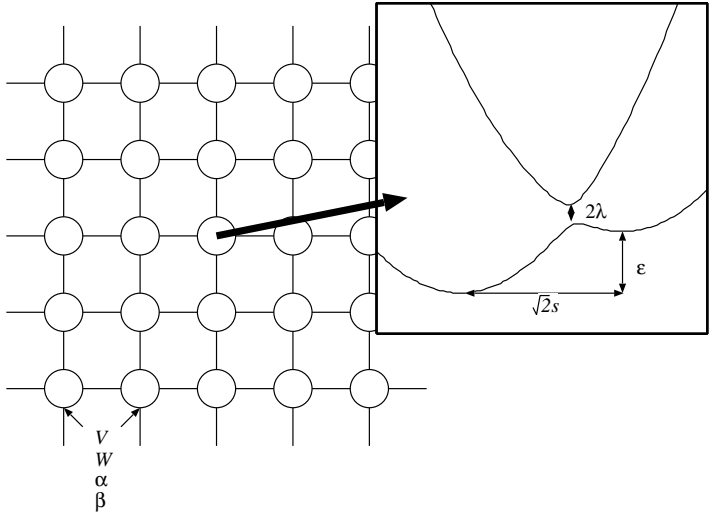}}
\caption{Schematic view of the model. Circles denote the molecules
  with two electronic states and a vibrational mode. Adiabatic potential energy
  surfaces for an individual molecule are shown in the inset.}
\label{schematic}
\end{figure}

A schematic view of the present model is shown in Fig.\
\ref{schematic}.
This figure shows that the two diabatic PESs for an
individual molecule cross each other, and that the nonadiabatic
coupling constant $\lambda$ acts to separate them into two adiabatic
PESs.
The parameters for the intra/intermolecular interactions are also shown in
the figure.
We chose the values of the parameters as:
$\varepsilon=2.3$,$s=1.4$,$V=1.1$,$W=0.2$, $\alpha=0.1$, $\beta=0.2$, and $\lambda=0.2$.
Although those values are typical for organic molecules as for
electron-vibration coupling\cite{wp} and  the intermolecular Coulomb interaction\cite{mol},
the other parameters are not easy to determine their values either from
theoretical calculations or experimental results.
We only mention that the order of magnitude for the parameters is
estimated referring to those for typical organic materials.
The model and the notations of the parameters are schematically shown in
Fig.\ \ref{schematic}.

We mention that the basis set for the vibronic states is composed of the Fock states
shown in Ref.\ \cite{old}.
The quantized states on each diabatic PES of a single molecule
are the vibronic states $| n \sigma \rangle_{\vec{r}}$ ($n =
0,1,2,...$, $\sigma=\uparrow,\downarrow$) in the Fock
representation, where the coordinate of the molecule is labelled by
$\vec{r}$.
$|n \uparrow \rangle$ is related with $|n
\downarrow \rangle$ by
\begin{equation}
| n \uparrow \rangle = |\uparrow \rangle \langle \downarrow
|e^{s(a^\dagger + a)} | n \downarrow \rangle,
\end{equation}
where $e^{s(a^\dagger +a)}$ denotes the translation operator in the
vibration coordinate space\cite{wp2}.
We note that this Ising-like model is similar to the one to study the thermodynamical
properties of the Jahn-Teller effect\cite{jahn}.

We obtain the numerical solution of the time-dependent Schr\"odinger equation for the
Hamiltonian (\ref{ham}) by the Runge-Kutta method.
In solving the time-dependent Schr\"odinger equation, we applied a mean-field
approximation in which the contribution of the wavefunction at the
nearest neighbor sites is substituted by the average value with
respect to the wavefunction.
Details of the quantization procedure of the vibration mode and the
method of calculation is described in Refs.\ \cite{ishida1} and \cite{ishida3}.

\section{calculated results}
\label{results}

\subsection{nucleation dynamics of isolated clusters of excited molecules}
\label{results1}

In this paper we study the dynamical pattern formation which is
dependent on the initial spatial distribution of excitation energy, i.e., density fluctuation
of excited molecule distribution\cite{ishida2}.
To be more precise, the density fluctuation  affects the interaction
energy coming from the intermolecular interactions, and thus the
dynamics of the system, the trigger of the
proliferation of excited molecules in particular, is determined by it.
Hence, it is helpful to pursue the time-evolution of small
clusters of excited molecules as an elementary process
of the nucleation dynamics.
In this subsection we aim at the classification of the nucleation
dynamics with respect to the initial configurations of  excited
molecules.

\begin{figure}
\scalebox{0.42}{\includegraphics*{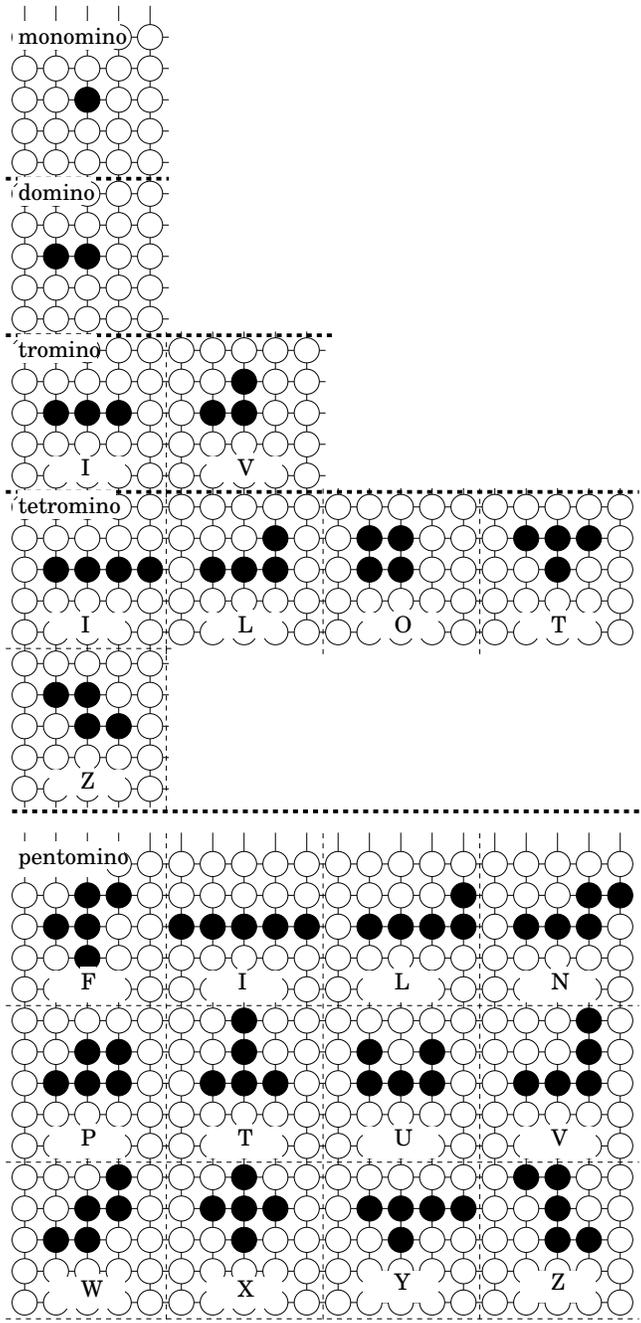}}
\caption{Cluster configuration of monomino, domino, trominoes,
  tetrominoes, and pentominoes.}
\label{clusters}
\end{figure}

Since we take into account the intermolecular interaction between
molecules at nearest neighbor sites, it is sufficient and adequate to
pursue the dynamics of ``connected molecular clusters'' in which all
the molecules of an individual cluster have others belonging to the same cluster in their adjacent sites.
In Fig.\ \ref{clusters} we show the configuration of the connected molecular clusters in which the number of 
excited molecules is up to five.

We have pointed out that the population of the excited electronic state $|\uparrow
\rangle_{\vec{r}}$  defined by
\begin{equation}
\tilde{N}(\vec{r},t) = \langle \Phi(t) | \hat{n}_{\vec{r}} | \Phi(t)
\rangle,
\end{equation}
where $|\Phi (t) \rangle $ denotes the solution of the time-dependent Schr\"odinger
equation,
makes the dynamical behavior of the initial nucleation processes visible\cite{ishida1,ishida2,ishida3}.
Besides, the sum of $\tilde{N}(\vec{r},t)$ over all the molecules
\begin{equation}
N(t) = \sum_{\vec{r}} \tilde{N}(\vec{r},t),
\end{equation}
is of interest since it is reminiscent of the non-conserved order parameter of 
the system in the terminology of nonequilibrium critical phenomena\cite{dcp}.
In this subsection we discuss the temporal behavior of $N(t)$
for the various types of the initial cluster.

\begin{figure}
\scalebox{0.5}{\includegraphics*{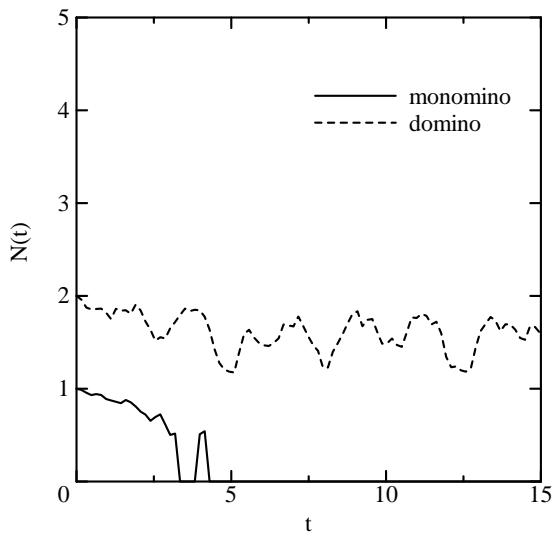}}
\caption{Population of the excited electronic state $N(t)$ on
  $64 \times 64$ lattice. The solid line and the dotted line
  correspond to the time evolution of $N(t)$ starting with a monomino
  and a domino, respectively.}
\label{monodo}
\end{figure}

With the parameter values shown in the last section, we observe that  no nucleation
process is induced in the system for either single excitation or
double excitation.
Figure \ref{monodo} shows the calculate results of $N(t)$, where
a monomino or a domino of excited molecules is placed at $t=0$.
In both cases $N(t)$ does not increase but
rather decrease as relaxation process proceeds.
The only difference between these two cases is that an isolated domino is
preserved during the relaxation process, while a monomino of excited state molecule
releases the excitation energy to the vibration mode and returns to
the electronic ground state 
within several periods of molecular vibration.
Thus, although the proliferation of excited-state molecules is not
realized in these cases, the intermolecular interaction tends to keep
the molecules on nearest neighbor sites in the excited
state $|\uparrow \rangle_{\vec{r}}$.

When three molecules which  form a tromino are excited to their Franck-Condon state at $t=0$
the dynamical behavior of the order parameter $N(t)$ is
qualitatively different from the above cases.

\begin{figure}
\scalebox{0.5}{\includegraphics*{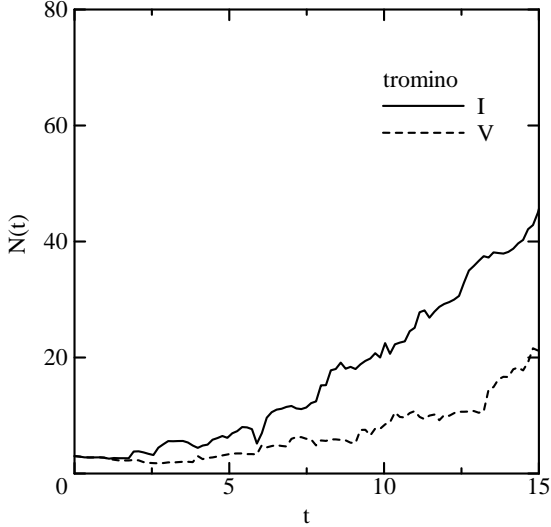}}
\caption{Population of the excited electronic state $N(t)$ on
  $64 \times 64$ lattice. The solid line and the dotted line
  correspond to the time evolution of $N(t)$ starting with a monomino
  and a domino, respectively.}
\label{tromino}
\end{figure}
Figure \ref{tromino} shows $N(t)$ when a tromino of excited molecules
are initially placed in the system.
The initial configurations of the excited molecules corresponding to each line are shown in
Fig.\ \ref{clusters}.
Contrary to the cases of monomino and dominoes, the number of excited molecules increases
in these cases as the relaxation process proceeds, which shows that
photoinduced nucleation process is triggered by a tromino of excited
molecules.
However, the (average) increase rate of $N(t)$ is different between
an I-tromino and a V-tromino, which is inferred as an importance of
the many-body correlation between excited molecules.
\begin{figure}[htbp]
\scalebox{0.5}{\includegraphics*{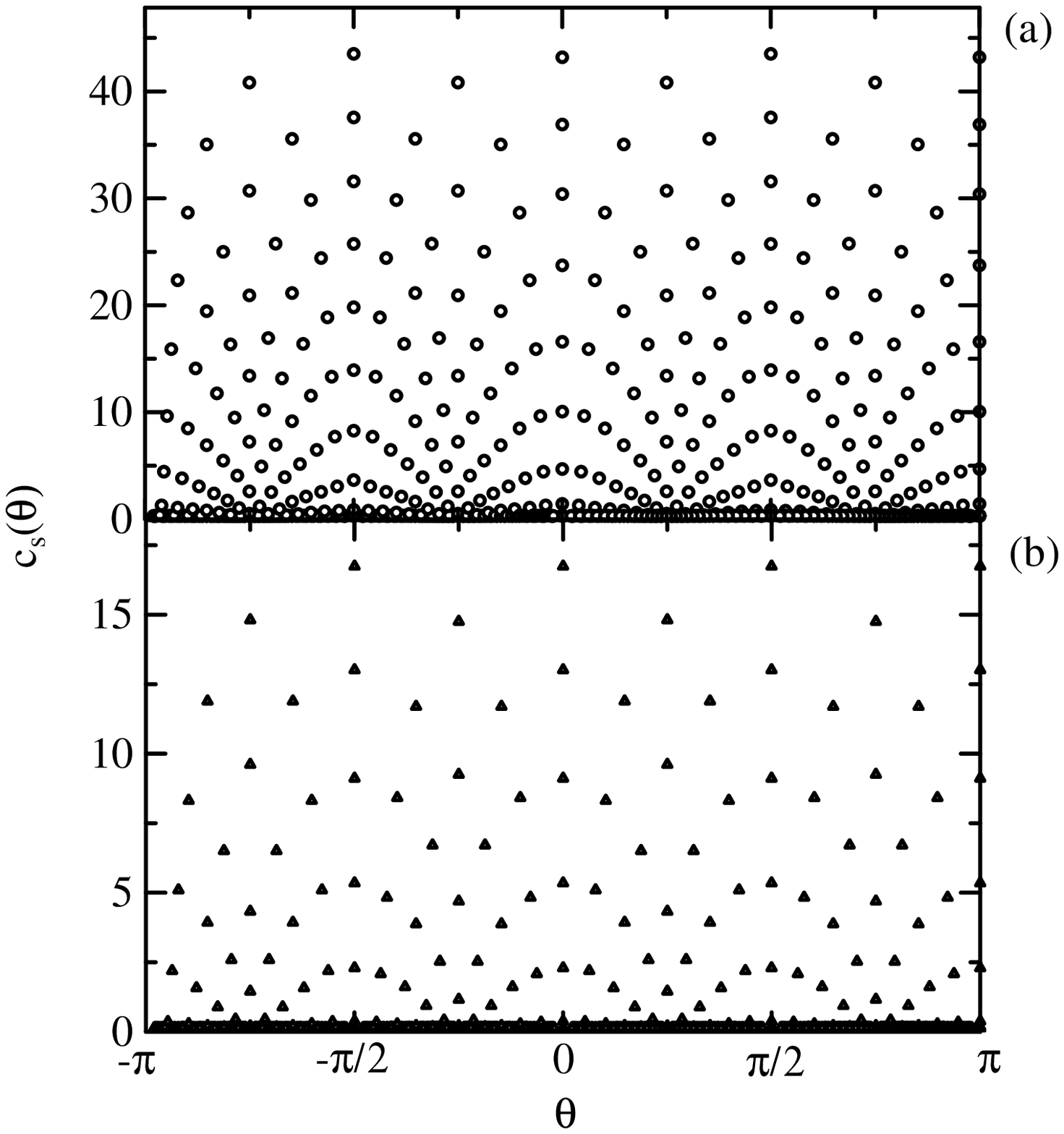}}
\caption{The correlation functions $c_s(r,\theta,t)$ for $t=15T$ are shown as a function of $\theta$.
Initial configurations of the excited-molecule clusters are: (a) an I-tromino and (b) a V-tromino.}
\label{singlecor}
\end{figure}
To understand the difference between the two cases, we first calculated the two-point
correlation function
\begin{equation}
c_s (r,\theta,t) = \sum_{\vec{r'}} \langle \Phi (t) | \hat{n}_{\vec{r'}} \hat{n}_{\vec{r}+\vec{r'}} | \Phi (t) \rangle,
\end{equation}
where $r$ and $\theta$ denote the radial and the angular component of $\vec{r}$, i.e.,
$\vec{r}=(r \cos \theta, r\sin \theta)$.

Figures \ref{singlecor}-(a) and (b) show  $c_s(r,\theta,t)$ for $t=15T$ as functions
of $\theta$.
The initial configuration of excited-molecule clusters for each
figure is an I-tromino or a V-tromino, respectively.
Since we are interested in a single domain, $c_s(r,\theta,t)$ has
larger value for smaller value of $r$ in general.
We found that, although the Hamiltonian (\ref{ham}) has the symmetry
of a square lattice $D_{4h}$, Fig.\ \ref{singlecor}-(a) shows that the 
growing domain of the excited molecules is not symmetric under $\pi/2$
rotation around the axis perpendicular to the lattice. 
To be more precise, the growing domain is symmetric under the $\pi$-rotation around the same axis
which is understood by the symmetric property of the I-tromino.
On the other hand, when a V-tromino is placed at $t=0$, the growing domain seems to be symmetric under 
the $\pi/2$ rotation around the same axis.
Although a V-tromino has lower symmetry than an I-tromino, we observe
that the V-tromino turns to be an O-tetromino at the early stage of the
relaxation process, and thus the symmetry of the cluster becomes higher.
We note that these asymmetric properties in $c_s(r,\theta,t)$ are
observed particularly for larger value of $r$, which shows that the geometrical structure at the perimeter of the
domain is strongly affected by the symmetry of the initial cluster.

The dynamical property of the nucleation process in the case of trominoes is understood based on
the above discussion.
Although we quantized the vibration modes instead of considering
 classical potential energy surfaces (PESs), it is
still feasible to regard the relaxation process as a motion of a massive
point on certain PESs except in the vicinity of the crossing points of diabatic PESs.
When a cluster of initially excited molecules has the same symmetry as the Hamiltonian,
the PES on which the relaxation process proceeds is also invariant 
under the symmetrical operations of the lattice.
In this case molecules in symmetric positions suffer the transition between $| \downarrow \rangle_{\vec{r}}$ and $| \uparrow \rangle_{\vec{r}}$ simultaneously.
Since the electronic state transition takes place by overcoming potential energy barrier between them, the required energy for the transition is larger as the symmetry is higher.
On the contrary, when the symmetry of the system is lowered by the
configuration of the initial cluster,
the degeneracy of the paths is released and particular paths which are
energetically favorable become to be chosen.
This means that the relaxation process accompanied by successive excitation of other molecules proceeds
more rapidly, and the growth of the I-tromino is faster than that of the V-tromino.

\begin{figure}
\scalebox{0.5}{\includegraphics*{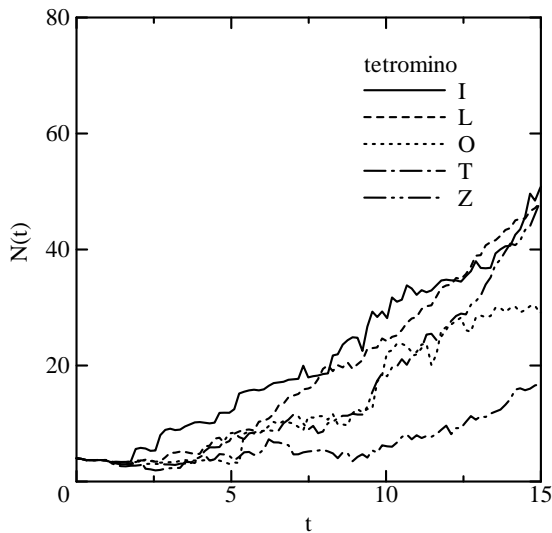}}
\caption{Population of the excited electronic state $N(t)$ on
  $64 \times 64$ lattice. The solid line and the dotted line
  correspond to the time evolution of $N(t)$ starting with a monomino
  and a domino, respectively.}
\label{tetromino}
\end{figure}

In the case of tetrominoes of excited molecules initially placed in the
system, we calculated five different configurations shown in Fig.\ \ref{clusters}.
As shown if Fig.\ \ref{tetromino}, the behavior of $N(t)$ for each
configuration shows rather different dynamical properties.
Proliferation of excited molecules starts earlier in an I-tetromino than
the other configurations, which shows that the symmetrical properties
of the electronic states discussed for trominoes affects the
time-dependence of $N(t)$.

\begin{figure}
\scalebox{0.42}{\includegraphics*{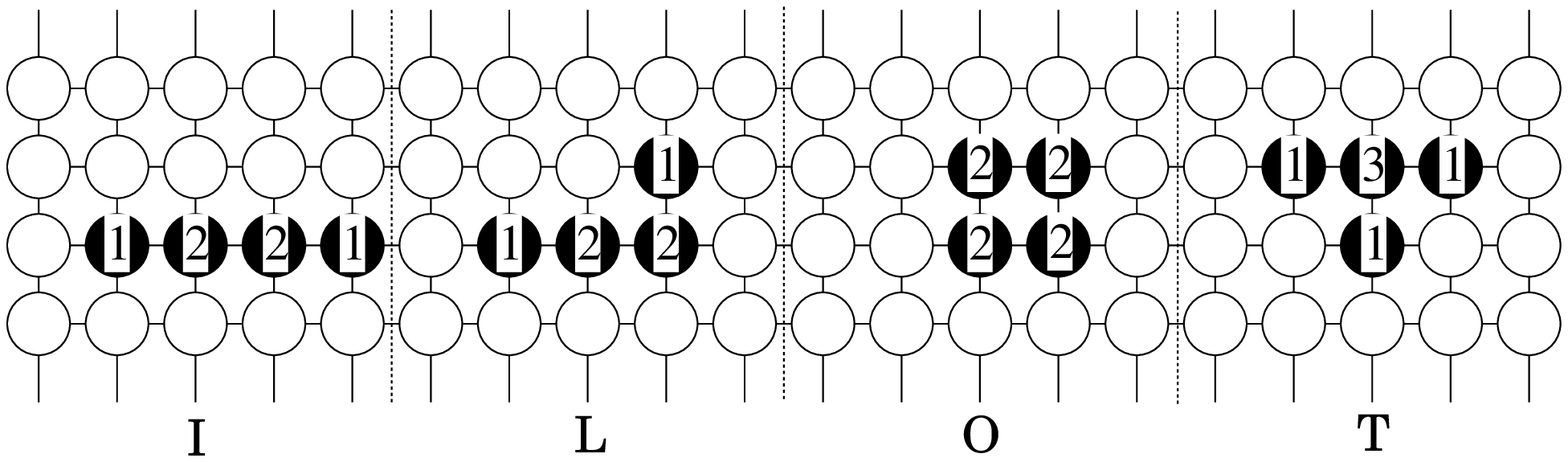}}
\caption{Number of neighboring molecules in tetrominoes. The black
  circles denote the excited molecule and the number in them shows the
number of neighboring black circles.}
\label{tetronbgh}
\end{figure}

Amongst the other configurations, the T-tetromino is slower than the other cases
in the domain growth process.
At the very initial stage of relaxation process, each excited molecule
begins to vibrate and adiabatic transition between $|\uparrow \rangle_{\vec{r}}$
and  $|\downarrow \rangle_{\vec{r}}$ takes place to make the molecule
 go back to the ground state.
When, however, some of its neighboring molecules are also in the excited
state, intermolecular interaction shown in Eq.\ (\ref{ham}) works to
keep the molecules in the excited state, and they become a
nucleus of the growing domain.
Thus, it is important to count the number of excited molecules in the
neighboring molecules to estimate the tendency of the cluster to be a
nucleus of the growing domain.
In the case of tetrominoes, the number of neighboring molecules is
shown in Fig.\ \ref{tetronbgh} for L-, O-, T-, and Z-tetrominoes.
We found in Fig.\ \ref{tetronbgh} that only T-tetromino has three molecules
with a single neighboring molecule, which shows that the T-tetromino
is easier to return to the ground state through the relaxation process than
the other tetrominoes.
However, Fig.\ \ref{tetromino} shows that the T-tetromino is
sufficient to trigger the nucleation process, although the growing
rate is lower than the other tetrominoes.

\begin{figure}
\scalebox{0.5}{\includegraphics*{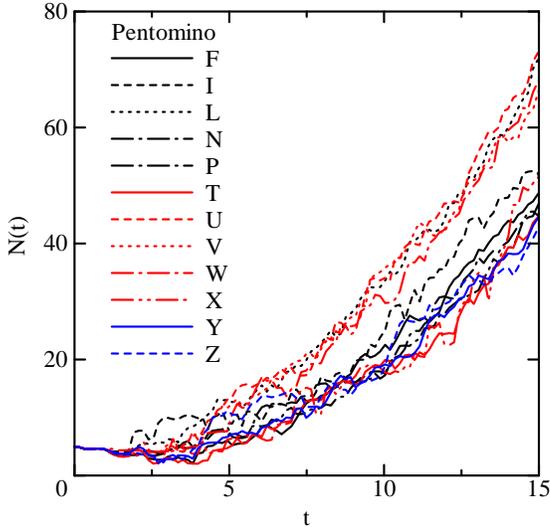}}
\caption{Population of the excited electronic state $N(t)$ on
  $64 \times 64$ lattice. The solid line and the dotted line
  correspond to the time evolution of $N(t)$ starting with a monomino
  and a domino, respectively.}
\label{pentomino}
\end{figure}

When a pentomino of excited molecules are placed initially, all of the
configurations are given sufficient excitation energy to be a nucleus of a growing domain as shown in Fig.\ \ref{pentomino}.
Though the behavior of $N(t)$ is quantitatively different for
each configuration, these differences are explained within the above
mechanisms, i.e., the spatial symmetry of the cluster and the number
of neighboring excited molecules at $t=0$.
Since, however, initial excitation energy for pentominoes is higher
than that for trominoes or tetrominoes, the nucleation is easier in
these cases, and thus the nucleation starts more smoothly in the case
of pentominoes.

We discussed in the previous paper\cite{ishida1} that the radius of the photoinduced domain
behaves as $\sim t^{1.2}$, which is understood by the picture that  
the growth of the domain is predominantly driven by propagation of coherent phonons 
rather than diffusion processes.
Figs.\ \ref{monodo}, \ref{tromino}, \ref{tetromino}, and
\ref{pentomino} shows that this feature is maintained in the present
case, and thus the coherent motion of molecular distortion is
important.
Since vibrational coherence survives for a few picoseconds in typical organic
molecules\cite{wp}, the present calculation is valid only in the 
time range studied in this paper, and the decoherence of the vibrational
states should be taken into account to study the growth dynamics of the
photoinduced domain in a longer time scale.

\subsection{pattern formation dynamics of randomly distributed excited molecules}

The calculated results presented so far show that the spatial distribution of the
initially excited molecules strongly affects their dynamics.
In other words, the density fluctuation of molecules in their excited state
is crucial to determine the destiny of each molecule, and thus is
important in the pattern formation dynamics of photoinduced nucleation.

When an aggregate of interacting molecules is irradiated by a laser pulse, a
certain amount of the constituent molecules absorbs photons.
We assume that the excitation strength $\rho$, the ratio of the number of the excited molecules to the
total number of molecule,
is constant determined by the light intensity and the absorption coefficient of the
molecules.
We note that there are numerous configurations of excited
molecules for a fixed value of $\rho$, and the quantum mechanical states
of the molecules are their superposition after the photoirradiation.
When we neglect the interference between these states after time evolution,
which is the case for small value of $\rho$ in particular, it is feasible to treat them independently.
In this subsection we discuss the dynamics of excited molecules when
multiple molecules are initially excited by a laser pulse, for example.
To calculate the time evolution of multiple excited molecules,
randomly selected  molecules on 96$\times$96 lattice with a periodic boundary condition is initially in the
Franck-Condon state, while the others are in the ground state.
For each value of $\rho$, we fix the number of initially excited molecules and calculated a
series of simulations with 64 different configurations of excited molecules, and the average values
over these series are obtained.

\begin{figure}
\scalebox{0.7}{\includegraphics*{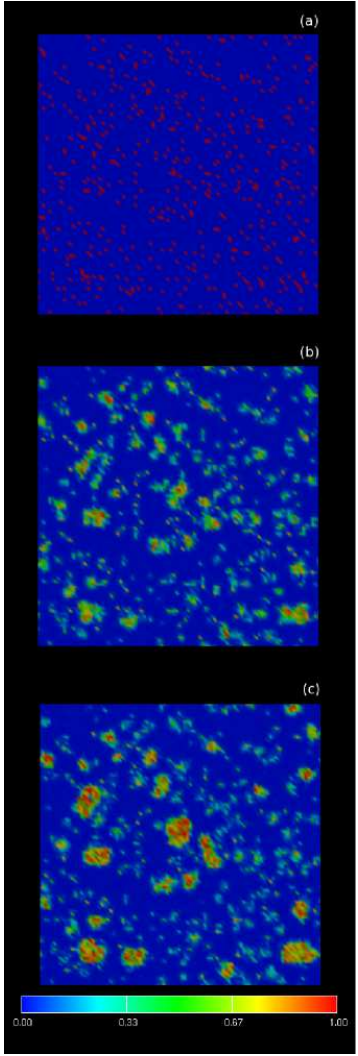}}
\caption{Gradation maps of excited state population $\tilde{N}(\vec{r},t)$on a 96$\times$96 lattice for
  $\rho=0.05$ for (a)$t=0$, (b) $t=7.5T$, and (c) $t=15T$. }
\label{480}
\end{figure}

\begin{figure}
\scalebox{0.7}{\includegraphics*{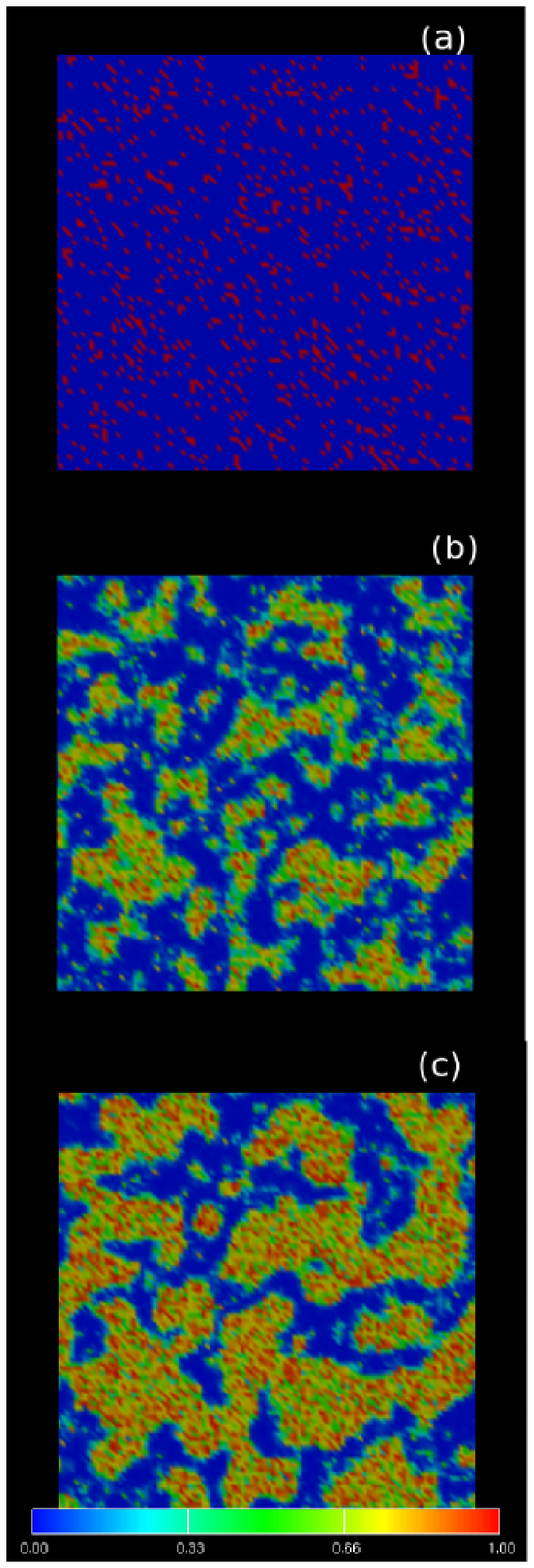}}
\caption{Gradation maps of excited state population $\tilde{N}(\vec{r},t)$on a 96$\times$96 lattice for
  $\rho=0.1$ for (a)$t=0$, (b) $t=7.5T$, and (c) $t=15T$. }
\label{960}
\end{figure}

Figures \ref{480} and \ref{960} show the snapshots of $\tilde{N}(\vec{r},t)$ 
for $\rho = 0.05$ and 0.1.
At $t=0$ the distribution of the excited molecules is not 
uniform, but the fluctuation of $\tilde{N}(\vec{r},t)$ is present.
When excited molecules are densely concentrated in certain parts of
the system, the molecules around them are able to overcome
the potential energy barrier to make the electronic state conversion, 
and thus the nucleation process is triggered there.
On the contrary, when the density of the excited molecules is
not sufficiently high to enable the nucleation process, the excitation energy is released
to the other molecules through the vibrational coupling $\alpha$, and
thus the excited molecules return to the ground state.
As a result, we obtain islands of photoinduced domains shown in Figs.\
\ref{480} and \ref{960} around densely distributed excited molecules.
When the distance between the islands is not large, they merge with
each other at the next step of the domain growth process to make larger ones as
shown in Fig.\ \ref{960}.
Apparently the section of the area of excited molecules is
nonlinearly dependent on $\rho$ as Figs. \ref{480} and \ref{960}, and
the density fluctuation of the initially excited
molecules also reflects the structure of the photoinduced domains,
and {\it vice versa}.
We note that these results directly reflects the discussion on the
smallest cluster for the domain growth shown in Sec.\ \ref{results1}.

We also calculated the following quantities which reveal the various
aspects of dynamical properties of the quantum pattern formation.
As we mentioned before, the average values over 64 independent series of simulations with different
initial conditions are discussed, which are denoted by $\langle ..\rangle$ in the following equations.
\begin{enumerate}
\item Conversion ratio $c(\rho)$ defined by
\begin{equation}
c(\rho) = \frac{\langle N(t=15T)\rangle }{M},
\end{equation}
where $M$ is the total number of the molecules.
\item Nucleation rate $r(\rho,t)$ defined by
\begin{equation}
r(t;\rho) = \left \langle \frac{d}{dt}\log N(t) \right \rangle =
\left \langle N(t)^{-1}\frac{dN(t)}{dt} \right \rangle.
\end{equation}
\item Two-point correlation function $C(|\vec{r}|,t; \rho)$ defined by
\begin{eqnarray}
C(|\vec{r}|,t; \rho) & = & \sum_{\vec{r'}}\left (  \langle
\tilde{N}(\vec{r}+\vec{r'},t)\tilde{N}(\vec{r'},t) \rangle \right . \nonumber \\
& - & \left . \langle \tilde{N}(\vec{r}+\vec{r'},t)\rangle \langle \tilde{N}(\vec{r'},t) \rangle \right ).
\end{eqnarray}
Contrary to the similar correlation function in the case of a
single-domain calculation ($c_s(r,\theta,t)$),
the $\theta$-dependence in the present case is not relevant when average over the series of simulations
is taken. Hence, the correlation function is a function of the radial part and the time in this case.
\item Autocorrelation function $A(t_w,\tau;\rho)$ defined by
\begin{eqnarray}
A(t_w,\tau; \rho) & = & {1\over M}\sum_{\vec{r}}\left (  \langle
\tilde{N}(\vec{r},t_w)\tilde{N}(\vec{r},t_w+\tau) \rangle  \right
. \nonumber \\
& - & \left . \langle
\tilde{N}(\vec{r},t_w)\rangle \langle \tilde{N}(\vec{r},t_w+\tau) \rangle \right ),
\end{eqnarray}
which is the temporal correlation between the fluctuation of excited-state population.
$M$ is the number of molecules and is 9216(=96$^2$) in the present calculations.
We note that, since we are interested in nonequilibrium dynamics of photoinduced
nucleation process, the autocorrelation function is a function of both
$t_w$ and $\tau$.
\end{enumerate}

\begin{figure}[htb]
\scalebox{0.45}{\includegraphics*{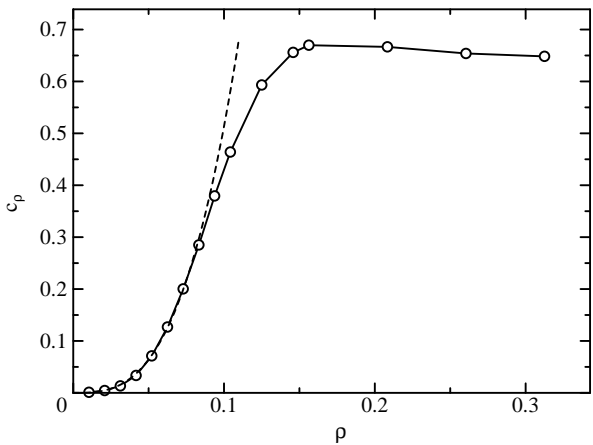}}
\caption{Conversion rate $c_\rho$ as a function of excitation ratio $\rho$. The dashed line which is proportional to $\rho^3$ is drawn as a guide for the eyes.}
\label{conv}
\end{figure}

We found that the size of the photoinduced domains nonlinearly depends
on the excitation ratio $\rho$.
Figure \ref{conv} shows the conversion ratio $c_\rho$ defined by
$c_\rho=N(t=15T)/M$ as a function of $\rho$.
To obtain these results we calculated the average value of $c_\rho$ for
each value of $\rho$ over 64 series of simulations as in the
calculation of $R$.

Figure \ref{conv} shows that $c_\rho$ depends on $\rho$ as $\sim \rho^3$ in the
dilute limit ($\rho \sim 0$), and deviates from $\rho^3$ for $\rho > 0.1$.
This feature reflects the size of the smallest cluster which enables
the growth of photoinduced domains.
With a fixed value of $\rho$, the smallest clusters for domain growth (a tromino) 
appear in the initial state with a probability proportional to $\rho^3$ in the dilute limit.
Hence, only a portion  of the initially excited molecules contributes
to the domain growth.
As we can neglect the interference between domains for $\rho \sim 0$, the
number of converted molecules is proportional to $\rho^3$ in this case.

As $\rho$ increases, the growing domains interfere with each other
and the domain growth becomes slower.
Thus the increase in $c_\rho$ is also slower than $\rho^3$ as the value of $\rho$ increases, which is
shown in Fig.\ \ref{conv}.
In any case, the conversion ratio increases as $\rho^m$ where $m$ is the
size of the smallest cluster which triggers the nucleation processes.
If $m$ is experimentally determined through the measurement of optical
properties, for example, we will have a clue to understand the microscopic
mechanism of the elementary process of the domain growth dynamics.

Figure \ref{conv} also shows that $c_\rho/\rho < 7$ in the present case, although some larger values were reported in experimental studies\cite{ct,pda,spin}.
Larger values of $c_\rho/\rho$ were found in materials close to their
critical temperature, since instability of thermodynamical state of the system enhances the
conversion ratio.
Hence, we mention that the present results correspond to
the cases away from  the critical temperature.
Furthermore, we should note that the present calculations are valid before the decoherence of vibrational states takes place,
and that the value of $c_\rho$ shown in Fig.\ \ref{conv} corresponds to that for $t \sim 3$ psec for $T \sim 200$fsec as in typical organic molecules.
We, however, expect that the domains continue to grow after the decoherence occurs,
and thus the experimentally obtained conversion ratio cannot be directly
compared with the present results quantitatively.
We stress that the nonlinearity of conversion ratio as a function of $\rho$ is essentially understood by the initial process of the domain growth,
and that the present calculations are of importance in order to understand the
microscopic mechanism of the photoinduced cooperativity.

\begin{figure}[htb]
\scalebox{0.45}{\includegraphics*{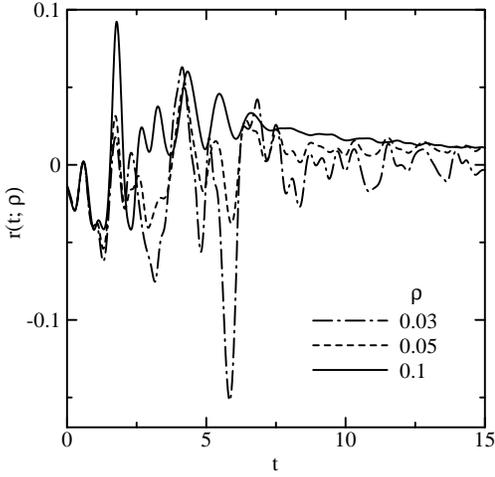}}
\caption{Nucleation rate $r(t;\rho)$ as a function of $t$.}
\label{grate}
\end{figure}
The average growth rate of the islands of excited molecules reveals another characteristic property of the dynamics of the initial nucleation process.
Figure \ref{grate} shows $r(t;\rho)$ for $\rho=0.03$, 0.05, and 0.1.
All the lines show that the oscillation of the nucleation rate is observed for $t
\stackrel{<}{\sim} 5T$ in accordance with the vibration of individual molecules.
$N(t)$ does not increase over this period, which shows that the nucleation process is not triggered.
At this stage of the relaxation process each cluster of excited molecules hesitates to grow as
also shown in Figs.\ \ref{tromino}, \ref{tetromino}, and \ref{pentomino}.
However, after such an ``incubation period'', $r(t; \rho)$ behaves in different ways according as $\rho$ is 
varied.
For $\rho=0.1$ the oscillation of the growth rate ceases and it is stabilized at a positive finite value,
which shows that that the domain growth rapidly progresses merging the islands and the domains of excited molecules 
cover the whole system.
On the contrary, for $\rho =0.03$ or $0.05$, the oscillation of $r(t; \rho)$ remains for $t>5T$, and $r(t;\rho)$ has 
a lowest (negative) value at $t\sim 6$ in both cases.
This means that the excited molecules start to turn back to the ground state, and that
 the nucleation process is not triggered after all.
For $\rho=0.05$, $r(t;\rho)$ decreases and thus the prominent domain growth still does not take place as shown in 
Fig.\ \ref{conv}.
In summary, $r(t; \rho)$ shows that the nucleation process is triggered at $t\sim 5T$ in the present case, and the 
domain growth starts only for sufficiently large values of $\rho$.

\begin{figure}[htb]
\scalebox{0.5}{\includegraphics*{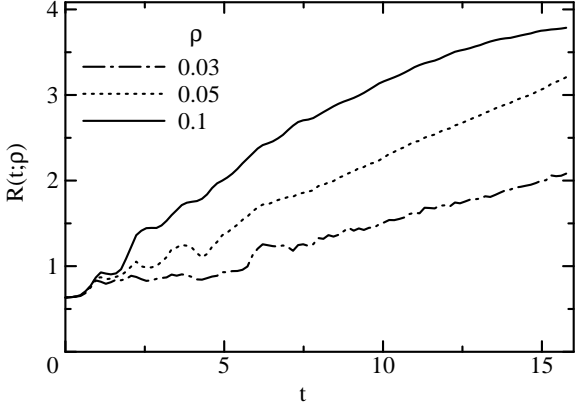}}
\caption{Correlation length $R(t;\rho)$ in units of the lattice constant for $\rho=0.03$,
  0.05, and 0.1 as functions of time.}
\label{radius}
\end{figure}

The two-point correlation function $C(|\vec{r}|,t; \rho)$ shows the
growth of the islands shown in Figs.\ \ref{480} and \ref{960}.
We define $R$ by the relation $C(R(t;\rho),t;\rho) = C(0,t;\rho)/e$, and $R(t;\rho)$ in units of the lattice constant
is shown in Fig.\ \ref{radius} for $\rho = 0.03$, 0.05, and 0.1.
All of the three lines in Fig.\ \ref{radius} show that $R(t;\rho)$ starts to increase
in the same manner, since $R(t;\rho)$ for $t \sim 0$ is determined by the
individual clusters of excited molecules.
Although Fig.\ \ref{radius} shows that $R(t;\rho)$ is proportional to $t$ for 
small values of $r$ as in the case of a single domain\cite{ishida1}, $R(t;\rho)$ increases more slowly for $\rho =0.1$.
This difference is due to the interference between growing domains, i.e., a domain disturbs the growth of the other ones
after they share perimeters with each other.
Hence, the increase rate of $R(t;\rho)$ slows down and it seems to behave as
$t^\alpha$ for $\alpha <1$ as in the diffusive domain growth\cite{ising,dcp}, for example.
However, we stress that the dynamics considered in the present study is always in coherent regime, and thus the slowdown of the growth rate 
is, as it were,  ``false'' $t$-dependence of $R(t;\rho)$.
Such a behavior of $R(t;\rho)$ should be discriminated from the diffusive domain growth in the systems which belong to a different universality class.
We mention that, since these properties reflects on the structure factor which is the Fourier transformation of $C(|\vec{r}|,t)$,
 they can be distinguished from each other experimentally by varying the excitation ratio.

\begin{figure}[htb]
\scalebox{0.5}{\includegraphics*{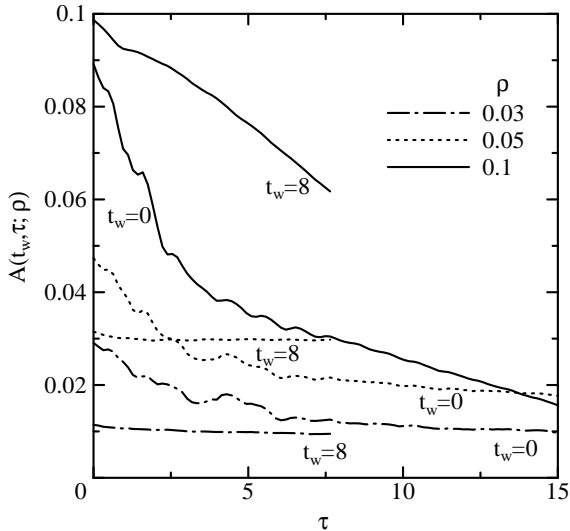}}
\caption{The autocorrelation function $A(t_w,\tau; \rho)$ for $\rho = 0.03$, 0.05, and 0.1.}
\label{acor}
\end{figure}

The autocorrelation function $A(t_w,\tau; \rho)$ also exemplifies the
dynamical aspects of the pattern formation processes.
Figure \ref{acor} shows the value of $A(t_w,\tau; \rho)$ as a function of $\tau$
for $\rho = 0.03$, 0.05, and 0.1.
For $t_w=0$ the autocorrelation function is described as 
\begin{equation}
A(0,\tau; \rho) = \frac{1}{M}\left \{\sum_{\rm excited} \langle \hat{N}(\vec{r},\tau)\rangle - \rho\langle N(\tau) \rangle \right \},
\label{acor0}
\end{equation}
where $\sum_{\rm excited} \cdots$ denotes the sum over initially excited molecules.
Thus, the value of $A(0,\tau;\rho)$ reflects the behavior of initially excited molecules, which will
converge on 0 when the growing domains cover the whole system.
In other words, as the excited-state domains extend over the system, $A(0,\tau; \rho)$ decreases, i.e., 
the temporal fluctuation of the excited-state population is suppressed.
However, when $\rho$ is small, the domains do not cover the whole system, and thus $N(\vec{r},\tau)$
depends on the initial configuration of the excited molecules. 
Hence, the temporal fluctuation remains finite as time goes by.
Therefore, $A(0,\tau; \rho)$ decreases faster for larger value of $\rho$ as shown in the figure.
For $t_w=8$, since the domain growth process is triggered before $t_w$,  the initial variation of 
excited-molecule distribution does not affect on $A(t_w,\tau; \rho)$.
Hence, the excited-state population at each site is closely correlated with each other and 
temporal correlation behaves in a similar manner to that for $t_w=0$ as $\tau$ becomes larger.

As shown in Eq.\ (\ref{acor0}), $A(0,\tau; 0.1)$ for large $\tau$ is proportional to $N(\tau)$, and hence
it linearly decreases with $\tau$ until all the molecules are converted to the excited state$|\uparrow \rangle_{\vec{r}}$.

\section{summary and conclusions}
\label{summary}

In this paper we study the coherent pattern formation dynamics of photoinduced
nucleation processes in organic molecular systems.
We showed that there exists a smallest cluster of excited molecules which triggers 
the nucleation process by our model Hamiltonian (\ref{ham}).
In such clusters the electronic state conversion
from $|\downarrow \rangle_{\vec{r}}$ to $|\uparrow \rangle_{\vec{r}}$ takes place successively and
photoinduced domain grows.
As in the results in Ref.\ \cite{ishida1} the basic scenario of the initial photoinduced nucleation
processes is that the excitation energy is transferred by coherent phonons
and the size of the converted domain (diameter) is almost linearly increases as the
nucleation proceeds.
As the decoherence of vibrational states takes place,
excitation energy propagation in the system will be dominated by 
diffusion processes, and hence the growth rate will be $\propto t$
after all.
We mention that these properties will be reflected on the time-resolved
spectra of {\it e.g.}, reflectance, absorbance, or Raman scattering
intensity and that the ultrafast spectroscopy
will give a key to understand the coherent nature of the nucleation processes.

Quantum pattern formation is observed in the photoinduced nucleation process.
When the irradiated photons induce excitation energy fluctuation in the system,
there appear islands of photoinduced nucleus which grows from the smallest cluster of
excited molecules.
The growth mechanism of such spatio-temporal patterns is purely quantum mechanical and
nonadiabatic transition between two electronic states plays an important role.
To make our points clearer, we calculated several correlation functions as well as the dependence
of conversion ratio on the initial excitation density.
Since multiple (3 in the present case) excited molecules are necessary to trigger the 
nucleation process, nonlinearity of the conversion ratio $c(\rho)$ is observed as a function of the 
initial excitation density $\rho$.
In particular, $c(\rho)\sim \rho^3$ in the dilute limit ($\rho \sim 0$) reflecting the probability of
the formation of the smallest cluster.
This result corresponds to the experimental results of reflectivity in polydiacetylenes\cite{pda}, and we 
stress that our calculated results have a correspondence to real systems.

The other properties shown in the last section revealed various aspects of the dynamics of
quantum pattern formation process.
The nucleation rate $r(t;\rho)$ shows that the nucleation process is similar for various values of $\rho$
for the first several periods of molecular vibration.
Then the effect of nucleation becomes apparent and $r(t;\rho)$ becomes almost constant when
domain growth rapidly proceeds.
This results supports the above-mentioned mechanism of photoinduced nucleation process, i.e., coherent
phonons plays a key role in the pattern formation.
The typical size of the domains is estimated by the two-point correlation function $C(\vec{r},t;\rho)$.
We found that the correlation length $R$ increases linearly as a function of $t$.
However, for $\rho=0.1$, the islands of excited molecules merge with each other and the increase rate of
the typical domain size is lower.
The autocorrelation function $A(t_w,\tau;\rho)$ also shows an interesting property of the dynamics of
the quantum pattern formation.
When separated islands of excited molecules are left in the system, the fluctuation of excited-state population remains 
finite, and thus $A(t_w,\tau;\rho)$ is long-lived for $\rho < 0.05$.
On the contrary, when domains of excited molecules cover the whole system, the autocorrelation function
decays as $\tau$ becomes large, and thus the pattern formation dynamics is divided into two categories
with this respect.

As we mentioned in the previous papers\cite{ishida1}, the validity of the 
mean-field approximation is also of importance in the quantum pattern formation.
If we consider a Gaussian wavepacket on the PES of a molecule, we
understand that the fluctuation of $u_{\vec{r}}$  is larger for smaller value of $\omega$.
Hence, the effect of the fluctuation is large when the motion of the
wavepacket is slow and the transition between PESs also takes place slowly.
To be more precise, the value of  $\lambda/\hbar \omega$ is a measure to estimate the validity of the
approximation, {\it i.e.} as the value of $\lambda/\hbar \omega$
is large, the mean-field approximation will becomes worse.
In the present calculation $\lambda/\hbar \omega=0.2$, and we expect
that the mean-field approximation works well.
On the other hand, when multiple wavepackets simultaneously move on the PESs of
a single molecule, the fluctuation of $u_{\vec{r}}$ and
$\hat{n}_{\vec{r}}$ becomes large, which is the case when the
population transfer takes place repeatedly during the nucleation 
processes.
Thus, at least in the present case, the mean-field approximation well describes
the physics of the photoinduced nucleation process.

In the present paper, we assume that only a single relevant
vibration mode exists in each molecule.
However, the nonadiabatic transition within a single molecule is
strongly affected by the structure of the PESs.
When multiple vibration modes are coupled to the electronic states,
the dynamics of the nucleation process depends on the topological structure of the intersections of the PESs, e.g.,
existence of conical intersections. 
In any case, {\it ab initio} electronic-structure calculations of specific materials will be
necessary in order to discuss such material-dependent features of the nucleation
processes.
We, however, stress that the present results give the basic properties
of the nucleation dynamics in coherent regime and that the qualitative
feature of the quantum pattern formation in the case of photoinduced nucleation process 
is sufficiently discussed in this paper.

Once we understand the mechanism of the quantum pattern formation dynamics of photoinduced 
cooperativity, we expect to develop a control methods of the dynamics by outer field, {\it
  e.g.}, laser pulses.
For this purpose, we should estimate the effect of decoherence of the
quantum-mechanical states.
We point out that it is possible to take into account the decoherence by
embedding the system in a large 'reservoir' and by tracing out the
dynamical variables regarding the reservoir.
The detailed structures of the spatial patterns is blurred as a result, 
and the contrast of the patterns discussed in the present paper will become lower.

We also mention that the present results will give a perspective to
the future experimental studies on the coherent dynamics of photoinduced
structure change by time-resolved X-ray diffraction measurement.
Since strong coherent X-ray sources are under development\cite{xfel},
it will be possible to observe the dynamics of the nucleation process and 
pattern formation process in coherent regime with femtosecond resolution,
and the present results will be compared with them to understand the physics of 
photoinduced cooperativity.

{\large \bf acknowledgments}\\
One of the authors(K.I.) thanks K. Takaoka, H. Asai, and S. Nunoue for helpful advice. 
This work was supported by the Next Generation Super Computing Project,  
Nanoscience Program, MEXT, Japan, and the numerical calculations were carried out on the computers at the
Research Center for Computational Science, National Institutes of Natural
Sciences.

\end{document}